\documentclass[twocolumn]{aastex631}

\newcommand{\oi}{i_{0}}
\newcommand{\ogi}{(g-i)_{0}}

\newcommand{\mv}{\rm{M_{V}}}
\newcommand{\mg}{\rm{M_{g}}}

\newcommand{\mm}{(m-M)_{0}}
\newcommand{\gmh}{\rm{[M/H]}}
\newcommand{\feh}{\rm{[Fe/H]}}
\newcommand{\afe}{\rm{[\alpha/Fe]}}
\newcommand{\rr}{\rm{R_{25}}}

\newcommand{\rh}{\rm{R_{h}}}
\newcommand{\sqdeg}{\rm{deg^2}}
\newcommand{\msun}{\rm{M}_{\odot}}

\begin{document}
\title{Uncovering the Ghostly Remains of an Extremely Diffuse Satellite in the Remote Halo of NGC\,253 \footnote{This research is based on data collected at the Subaru Telescope, which is operated by the National Astronomical Observatory of Japan.}}

\author[0000-0002-7866-0514]{Sakurako Okamoto}
\affiliation{Subaru Telescope, National Astronomical Observatory of Japan,\\ 650 North A'ohoku Place, Hilo, HI 96720, U.S.A.}
\affiliation{National Astronomical Observatory of Japan, Osawa 2-21-1, Mitaka, Tokyo, 181-8588, JAPAN}
\affiliation{The Graduate University for Advanced Studies, SOKENDAI, Osawa 2-21-1, Mitaka, Tokyo 181-8588, JAPAN}

\author[0000-0001-7934-1278]{Annette M.N. Ferguson}
\affiliation{Institute for Astronomy, University of Edinburgh, Royal Observatory, Blackford Hill, Edinburgh, EH9 3HJ U.K.}

\author{Nobuo Arimoto}
\affiliation{National Astronomical Observatory of Japan, Osawa 2-21-1, Mitaka, Tokyo, 181-8588, JAPAN}
\affiliation{The Graduate University for Advanced Studies, SOKENDAI, Osawa 2-21-1, Mitaka, Tokyo 181-8588, JAPAN}

\author[0000-0001-8239-4549]{Itsuki Ogami}
\affiliation{The Graduate University for Advanced Studies, SOKENDAI, Osawa 2-21-1, Mitaka, Tokyo 181-8588, JAPAN}

\author[0000-0002-8566-0491]{Rokas \v{Z}emaitis}
\affiliation{Institute for Astronomy, University of Edinburgh, Royal Observatory, Blackford Hill, Edinburgh, EH9 3HJ U.K.}

\author[0000-0002-9053-860X]{Masashi Chiba}
\affiliation{Astronomical Institute, Tohoku University, 6-3 Aoba, Aramaki, Aoba-ku, Sendai, Miyagi 980-8578, JAPAN}

\author[0000-0002-2191-9038]{Mike J. Irwin}
\affiliation{Institute of Astronomy, University of Cambridge, Madingley Road, Cambridge CB3 0HA, U.K.}

\author[0000-0002-2502-0070]{In Sung Jang}
\affiliation{Department of Astronomy \& Astrophysics, University of Chicago, 5640 South Ellis Avenue, Chicago, IL 60637, U.S.A.}

\author[0000-0002-8762-7863]{Jin Koda}
\affiliation{Department of Physics and Astronomy, Stony Brook University, Stony Brook, NY 11794-3800,  U.S.A.}

\author[0000-0002-3852-6329]{Yutaka Komiyama}
\affiliation{Dept. of Advanced Sciences, Faculty of Science and Engineering, Hosei University, 3-7-2 Kajino-cho, Koganei-shi, Tokyo 184-8584, JAPAN}

\author[0000-0003-2713-6744]{Myung Gyoon Lee}
\affiliation{Astronomy Program, Department of Physics and Astronomy, SNUARC, Seoul National University, 1 Gwanak-ro, Gwanak-gu, Seoul 08826, Republic of Korea}

\author[0000-0003-3301-759X]{Jeong Hwan Lee}
\affiliation{The Center for High Energy Physics, Kyungpook National University, 80 Daehakro, Bukgu Daegu 41566, Republic of Korea}

\author[0000-0003-0427-8387]{Michael Rich}
\affiliation{Department of Physics and Astronomy, The University of California, Los Angeles, CA 90095,  U.S.A.}

\author[0000-0002-5011-5178]{Masayuki Tanaka}
\affiliation{National Astronomical Observatory of Japan, Osawa 2-21-1, Mitaka, Tokyo, 181-8588, JAPAN}
\affiliation{The Graduate University for Advanced Studies, SOKENDAI, Osawa 2-21-1, Mitaka, Tokyo 181-8588, JAPAN}

\author[0000-0003-2258-7044]{Mikito Tanaka}
\affiliation{Dept. of Advanced Sciences, Faculty of Science and Engineering, Hosei University, 3-7-2 Kajino-cho, Koganei-shi, Tokyo 184-8584, JAPAN}



\begin{abstract}
We present the discovery of NGC253-SNFC-dw1, a new satellite galaxy in the remote stellar halo of the Sculptor Group spiral,  NGC\,253.  The system was revealed using deep resolved star photometry obtained as part of the Subaru Near-Field Cosmology Survey that uses the Hyper Suprime-Cam on the Subaru Telescope. Although rather luminous ($\mv = -11.7 \pm 0.2$) and massive ($M_* \sim 1.25\times 10^7~\msun$), the system is one of the most diffuse satellites yet known, with a half-light radius of $\rh = 3.37 \pm 0.36$ kpc and an average surface brightness of $\sim 30.1$ mag arcmin$^{-2}$ within the $\rh$.  The colour-magnitude diagram shows a dominant old ($\sim 10$ Gyr) and metal-poor ($\gmh=-1.5 \pm 0.1$ dex) stellar population, as well as several candidate thermally-pulsing asymptotic giant branch stars. The distribution of red giant branch stars is asymmetrical and displays two elongated tidal extensions pointing towards NGC\,253, suggestive of a highly disrupted  system being observed at apocenter.  NGC253-SNFC-dw1 has a size comparable to that of the puzzling Local Group dwarfs Andromeda\,XIX and Antlia\,2 but is two magnitudes brighter.  While unambiguous evidence of tidal disruption in these systems has not yet been demonstrated, the morphology of NGC253-SNFC-dw1 clearly shows that this is a natural path to produce such diffuse and extended galaxies. The surprising discovery of this system in a previously well-searched region of the sky emphasizes the importance of surface brightness limiting depth in satellite searches. 
\end{abstract}




\section{Introduction}
The $\Lambda$CDM cosmological model predicts that galaxies form hierarchically; large galaxies like the Milky Way originate from small over-densities in the primordial matter distribution and grow via the agglomeration of numerous smaller building blocks, some of which survive later merging and represent the present-day dwarf satellites \citep[e.g.][]{2005ApJ...635..931B, 2014MNRAS.444..237P}.   The numbers and properties of satellite galaxies around massive hosts has emerged as a powerful way to test the  $\Lambda$CDM paradigm on small scales as well as the physics of galaxy formation \citep[e.g.][]{2017ARA&A..55..343B, 2022NatAs...6..897S}.  As a result, there has been much interest in characterising the satellite system of the Milky Way and its nearest neighbours  \citep[e.g.][]{2019PASJ...71...94H,2022ApJ...933...47C, 2023ApJ...952...72D}.  Although the sample size beyond the Local Group (LG) is still limited, such observations are critical to benchmark results from the Milky Way and M31. Recent wide-field imaging and/or survey campaigns are having remarkable success in uncovering new satellites around Local Volume galaxies \citep[e.g.][]{2013AJ....146..126C, 2016ApJ...823...19C, 2019ApJ...884..128O,2022ApJ...926...77M} but these studies rapidly lose sensitivity at surface brightnesses $\mu_V \gtrsim 28-29$~mag~arcsec$^{-2}$, raising questions about what they might be missing. 

NGC\,253, lying at D $\sim 3.5$ Mpc \citep{2011ApJS..195...18R},  is a barred spiral galaxy located in the Sculptor group.  It has a similar luminosity and virial radius to the Milky Way and M31 \citep{2013AJ....145..101K, 2021ApJ...918...88M}, making it a prime object for comparative studies with the LG spirals.  An extended and structured stellar halo has long been known in this galaxy, detected in both integrated light \citep[e.g.][]{1982A&A...106..112B, 1997PASA...14...52M} and resolved star counts \citep{2010ApJ...725.1342D, 2011ApJ...736...24B, 2014A&A...562A..73G,2017MNRAS.466.1491H}.  It also hosts a rich population of halo globular clusters \citep{2018A&A...611A..21C}.  

In recent years, attention has focused on the faint satellite system of NGC\,253. The PISCeS survey used Magellan/Megacam resolved star photometry to discover five new dwarf satellites of NGC\,253 \citep{2014ApJ...793L...7S, 2016ApJ...816L...5T, 2022ApJ...926...77M}, with two of them, Scl-MM-dw2 (hereafter MM-dw2) and Scl-MM-dw3, independently discovered as NGC253-dw2 by \citet{2016MNRAS.457L.103R} and Donatiello\,II by \citet{2021A&A...652A..48M}, respectively.  \citet{2021A&A...652A..48M} proposed two further dwarf galaxy candidates from Dark Energy Survey data and \citet{2022ApJ...933...47C} suggested another one from DECaLS data, both from integrated light analyses. \citet{2024arXiv240114457M} confirm these latter systems with Hubble Space Telescope photometry and undertake a comprehensive study of the NGC\,253 satellite system as a whole, finding it to be deficient in luminous satellites compared systems of similar stellar mass. 

In this Letter, we report the discovery of a highly-disrupted satellite galaxy in the outer halo of NGC\,253.  Although luminous, the system is extremely diffuse and thus has been missed by all previous searches in this part of the sky. The discovery represents the first result from the ``Subaru Near Field Cosmology'' survey (hereafter SNFC), a systematic imaging survey of a sample of late-type galaxies in the Local Volume in order to study their resolved stellar halos.  In Section 2, we summarize the observations and data reduction methodology.  We derive the physical properties of the dwarf,  which we name NGC\,253-SNFC-dw1 (hereafter SNFC-dw1), in Section 3 and discuss and conclude in Section 4.

\section{Observation and data} \label{sec:obs}
\begin{figure*}
\begin{center}
\includegraphics[width=510pt]{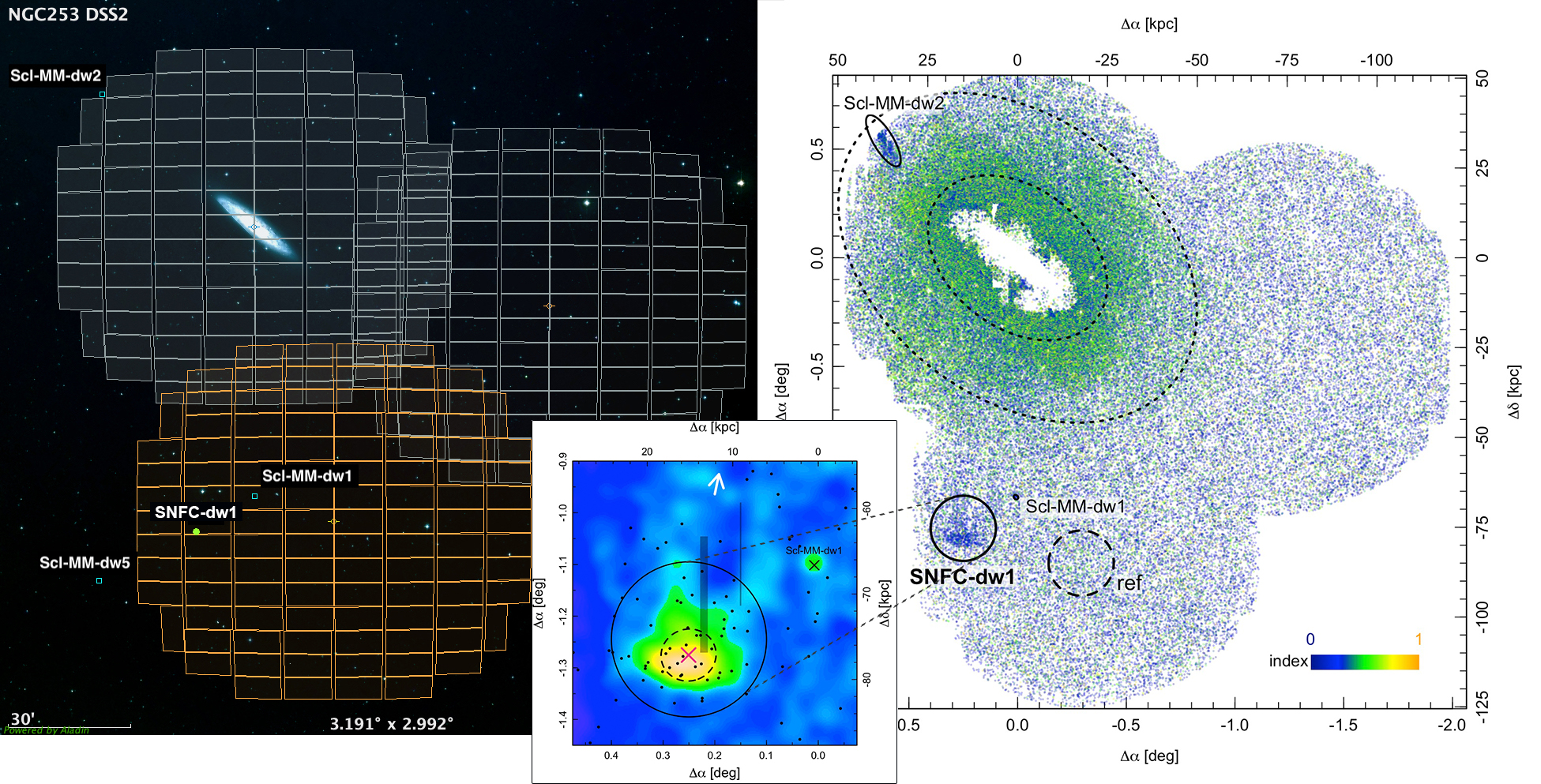}
\caption{{\it Left:} The HSC pointings around NGC\, 253 overlaid on the Digitized Sky Survey image using the Aladin Sky Atlas\citep{2000A&AS..143...33B}. The orange tile indicates the pointing where SNFC-dw1 is discovered. The location of SNFC-dw1 and the previously-known satellites within the footprint are indicated.  {\it Right:} The spatial distribution of RGB stars selected using the polygon in Figure \ref{fig: 4cmds}(a). The color of each point represents the $\ogi$ color of the star with transparency. The dotted ellipses indicate the $2\times\rr$ and $4\times\rr$ radii of NGC\,253 \citep{1991rc3..book.....D}, assuming b/a=0.7. The solid and dashed circles show the circular radius of 9 arcmin centered on SNFC-dw1 and a nearby reference field. The $2.5\times\rh$ of the two known dwarf satellites within the FOV are also shown as solid lines and labeled.  {\it Lower middle:} The smoothed density map of metal-poor RGB stars centered on SNFC-dw1 and selected within the left part of the solid polygon of Figure \ref{fig: 4cmds}(b).  The magenta cross shows the estimated centroid of the structure, of which the coordinate is shown in Table\ref{tbl: str}. The dashed line shows the fitted $\rh$ of the S\'{e}rsic profile. The black points represent the TP-AGB candidates. Gray-shaded areas are the masked regions affected by saturated foreground stars. The white arrow indicates the direction toward the NGC\,253 center.}
\label{fig: map}
\end{center}
\end{figure*}

We observed three pointings around NGC\,253 using the Hyper Suprime-Cam (HSC) on the Subaru 8.2m telescope (left panel of Figure \ref{fig: map}). The HSC comprises 104 CCD detectors and provides a field-of-view (FOV) of 1.76 $\sqdeg$ with a pixel scale of 0.17\arcsec \citep{2018PASJ...70S...1M}. The observations were obtained in queue-mode as part of the open-use SNFC intensive program during the period 2019-2022.  For NGC\,253, $g$- and $i$-band images were acquired under seeing conditions ranging from 0.7-1.0\arcsec.  For the two northern fields, we obtained a total exposure time of 12500~sec and 2750~sec in the $g$- and $i$-bands, respectively. In the southern field, we obtained the same exposure time in the $i$-band but the $g$-band exposures amounted to only 6000~sec; furthermore, the seeing was slightly worse in this field than in the northern fields.  Our exposure times are designed to provide sensitivity to a broad range of stellar metallicities, including the most metal-rich halo stars.

The data were processed using the reduction pipeline hscPipe 8.4 \citep{2018PASJ...70S...5B}, which is based on a software suite being developed for the Vera C. Rubin Observatory data \citep{2010SPIE.7740E..15A, 2017ASPC..512..279J, 2019ApJ...873..111I}.  Each frame was corrected for bias, dark, and flat-fielding, and then the sky was modeled internally and subtracted with an aggressive 32-pixel mesh. The frames in each filter were calibrated using Pan-STARRS1 catalog \citep{2013ApJS..205...20M, 2016arXiv161205560C}, then mosaicked and coadded. The final photometry is in the HSC filter system and in AB magnitudes. Forced photometry based on the point spread function (PSF), CModel, and Kron model was performed on the coadded images, with the $g$-band image providing the initial detections.  Extended sources were excluded by considering the PSF to CModel flux ratio, following \citet{2019ApJ...880..104P}. In particular, we classified an object as a point-source if the flux ratio is within $1\sigma$ of unity in the $i$-band.  The point-source detection limits of the coadded images were $g\sim27.9$ and $i\sim26.3$ with SNR$=5$ for the northern two fields and $g\sim27.5$ and $i\sim25.8$ with SNR$=5$ for the southern field.  These depths correspond to roughly 1.5 to 2 mag below the tip of red giant branch (TRGB) at the distance of NGC\,253 \citep{2011ApJS..195...18R}. 

The Galactic extinction corrections were taken from the \citet{1998ApJ...500..525S} reddening map using the coefficients from \citet{2011ApJ...737..103S}. We applied an extinction correction to each source individually according to its location, assuming a \citet{1999PASP..111...63F} reddening law with $\rm{R_V}=3.1$.  The mean reddening across the field is rather small, amounting to $E(B-V) \sim 0.0161$.

Artificial star tests were performed on the coadded images to estimate the accuracy and completeness of the photometric catalog. Artificial point sources were created using the PSF models generated by the pipeline and injected into the coadded images as described in \citet{2024arXiv240100668O}. The resulting images were processed in the same way as the original ones.  About 30,000 artificial stars were added to the images in intervals of 1.0 mag from 20 mag to 24 mag, and 0.5 mag from 24 mag to 28 mag, in both the $g$- and $i$-bands. We considered the artificial stars to be recovered if the PSF to CModel flux ratio of the detected source was within $1\sigma$ of unity at the input magnitude.  
The completeness was examined in elliptical annuli centered on NGC\,253, assuming an axis ratio $\rm{b}/\rm{a} = 0.7$ which is judged by eye to be a good match to the shape of the outer halo.    
At the radial distances of relevance for this paper (30-110 arcmin), the recovery fraction is almost constant for a given input magnitude. 
We interpolated the recovery fractions to produce a 2D histogram representing the completeness fraction as function of magnitude and color, and used this to correct for incompleteness where required.

\begin{figure*}
\begin{center}
\includegraphics[width=500pt]{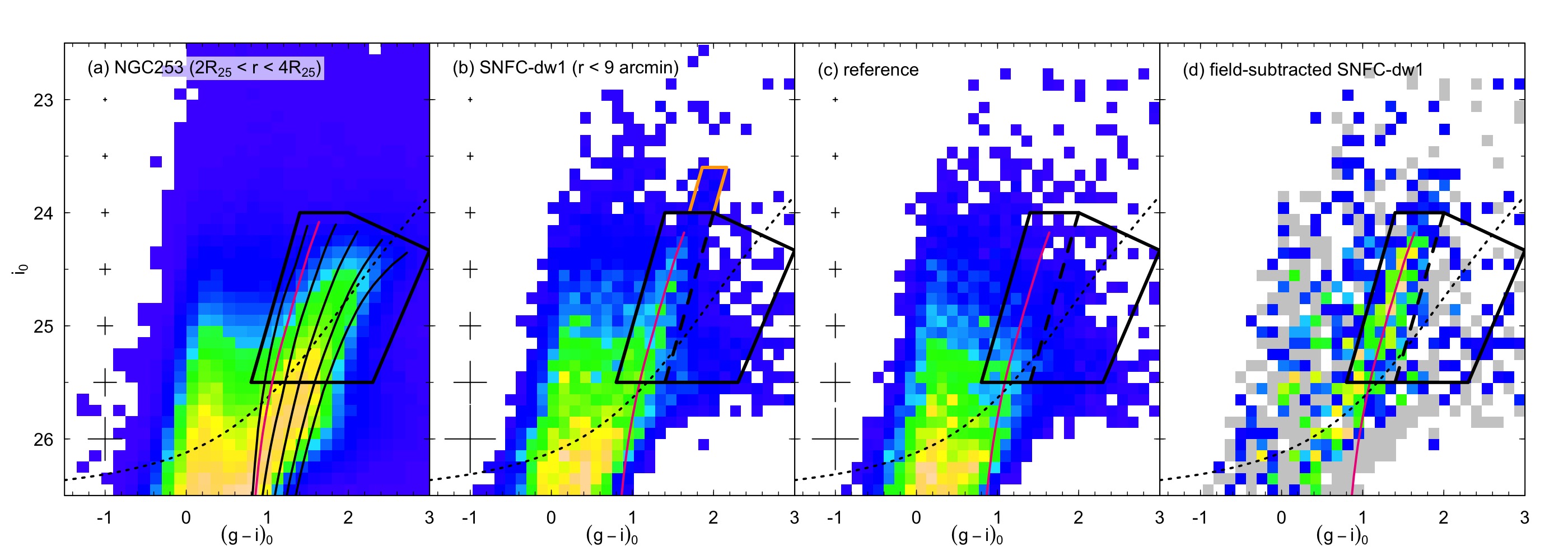}
\caption{De-reddened Hess diagrams of stellar objects in NGC\,253, the SNFC-dw1 region, a nearby reference field, and the field-subtracted SNFC-dw1 region.  The bin sizes are 0.15 mag in the x-axis and 0.1 mag in the y-axis. 
(a) Stars within an elliptical annulus of $2\times\rr < $ r $< 4\times\rr$ centered on NGC\,253, assuming b/a$=0.7$.  The solid polygon indicates the area used to select RGB stars for the right panel of Figure \ref{fig: map}. PARSEC isochrones of age 10 Gyr and $\gmh=-2.2, -1.8, -1.4, -1.0, -0.7, -0.4$ are shifted to NGC\,253's distance and overlaid as solid lines \citep{2011ApJS..195...18R,2012MNRAS.427..127B}.   
(b) Stars within a circular radius of 9 arcmin of SNFC-dw1. A 10 Gyr isochrone of $\gmh=-1.8$ is shifted to the estimated SNFC-dw1 distance and overlaid. The dashed line in the solid polygon is used to separate the metal-poor RGB stars. The orange solid parallelogram above the RGB selection box is used to select TP-AGB candidates.
(c) Stellar objects in a reference field, selected as an equal area region that lies at the same elliptical radius from NGC\,253 as SNFC-dw1. (d)  The field-subtracted Hess diagram of SNFC-dw1, created by subtracting panel (c) from (b).  The gray color indicates negative values.}
\label{fig: 4cmds}
\end{center}
\end{figure*}

\section{Properties of NGC253-SNFC-dw1} \label{sec:sub}

\subsection{Star Counts and Stellar Populations}

The right panel of Figure \ref{fig: map} shows the spatial distribution of RGB stars across three HSC pointings around NGC\,253. RGB stars are selected from stellar objects within the black solid polygon on the color-magnitude diagram (CMD) shown in Figure \ref{fig: 4cmds}(a), the boundaries of which have been chosen to optimize the selection of RGB stars at the distance of NGC\,253, while limiting the number of foreground and background contaminants. The color of each point in the map denotes the $\ogi$ color of the star, which can be interpreted as a rough proxy for metallicity.  We linearly convert the color and magnitude of each star to the index (0 to 1) and assign the color (dark blue to orange). The bluest (index$=0$) and reddest (index$=1$) colors correspond to $\gmh = -2.2$ and $-0.4$, respectively, assuming 10 Gyr old age \citep{2012MNRAS.427..127B}.  The map reveals a very extended and asymmetric metal-rich stellar halo that completely fills the central HSC pointing ($\textrm{R}_{proj} \sim 50$ kpc) and which will be the subject of a future paper.  

The map also reveals two very prominent metal-poor substructures at the edges of the current SNFC survey coverage. While the northern feature is the satellite galaxy, 
MM-Dw2, discovered independently by \citet{2016ApJ...816L...5T} and \citet{2016MNRAS.457L.103R}, the southern over-density, which we call SNFC-dw1, is new. Another known satellite, Scl-MM-dw1, is located near SNFC-dw1, but it is hardly visible on the map due to its compactness ($\rh = 16.8 \arcsec$) \citep{2014ApJ...793L...7S}.  To investigate the properties of SNFC-dw1, we first set the center by eye and isolate a surrounding circular region of 9 arcmin radius, as shown by the black solid circle. We also select a reference field of identical area at the same elliptical radius from NGC\,253's center as SNFC-dw1, shown by the black dashed circle.

The lower-middle panel of Figure \ref{fig: map} presents the zoom-in view around SNFC-dw1. It shows the smoothed density map of metal-poor RGB stars within the leftward of the dashed line in the polygonal region in Figure \ref{fig: 4cmds}(b). The kernel density is estimated with the bandwidth of 1.\arcmin0, which is $\sim 1$ kpc at the distance of NGC\,253. There are two areas affected by narrow diffraction spikes from saturated foreground stars where the pipeline failed to return reliable photometry. To reduce any spurious effects, we masked these regions,  shown with gray shading, and re-populated them with stars drawn from the adjacent area before estimating SNFC-dw1's properties.  The density map shows the flattened core of SNFC-dw1 with a surrounding asymmetrical extension.  Specifically, the stellar density rapidly decreases on the southern side, while on the northern side two elongated tidal extensions can be seen that point towards NGC\,253, indicated by the white arrow in the panel. This morphology indicates that SNFC-dw1 is highly-disrupted and that we are likely viewing it at apocenter,  with both the leading and trailing tails projecting on the same side of its main body.  

Figure \ref{fig: 4cmds} shows the de-reddened Hess diagrams of stellar sources around NGC\,253, SNFC-dw1, the reference field, and the field-subtracted SNFC-dw1. The error bars in panels (a) to (c) show the median photometric error of stars in the interval $-1.5 < \ogi < 3.0$.  The 50 percent completeness threshold is shown as the dotted line.  Panel (a) contains stars within an elliptical annulus of $2\times \rr < $ r $< 4\times\rr$ of NGC\,253, adopting the $\rr=13.7\arcmin$ \citep{1991rc3..book.....D}.  PARSEC isochrones of age 10 Gyr and $\gmh=-2.2, -1.8, -1.4, -1.0, -0.7, -0.4$ \citep{2012MNRAS.427..127B} are shifted to NGC\,253's distance and overlaid as solid lines.  A  well-populated broad RGB is seen  ($\oi > 24$, $\ogi >1$) that spans the full range of isochrones shown. Another obvious feature is the blue plume at $ -0.5 < \ogi < 0.7$, which we identify with unresolved high-z galaxies as seen in other wide-field surveys of nearby galaxies \citep[e.g.][]{2012MNRAS.419.1489B, 2019ApJ...884..128O} as well as in the reference field (Figure \ref{fig: 4cmds}(c)).  Comparing Figure \ref{fig: 4cmds}(b) that shows stars within a circular radius of 9 arcmin centered on SNFC-dw1  with Figure  \ref{fig: 4cmds}(c) that shows the equal area reference field,  the signature of a metal-poor RGB sequence is striking.  This feature is well-matched to a theoretical isochrone of $\gmh=-1.8$ and 10 Gyr old. A tantalising feature in panel (b) is the plume of stars that sits above the RGB selection box that we tentatively associate with a thermally-pulsing asymptotic giant branch (TP-AGB) population.  We select these stars using the orange solid parallelogram in panel (b), and plot their spatial distribution as the black points in the lower-middle panel of Figure \ref{fig: map}.  These sources are indeed visually clustered on the main body of SNFC-dw1 as would be expected for an instrinsic population.   The field-subtracted Hess diagram in Figure \ref{fig: 4cmds}(d) confirms the clear RGB overdensity as well as the existence of TP-AGB stars in the SNFC-dw1 field. 

\subsection{Distance}
We derive the distance to SNFC-dw1 using the TRGB method \citep[e.g.][]{1993ApJ...417..553L}, following the procedure described in \citet{2019ApJ...884..128O}. Stars of $1.2 < \ogi < 2.25$ that lie within 9 arcmin radius are used to derive the $i$-band luminosity function (LF).  Using the LF of point sources in the reference field, we correct for contamination from the NGC\,253 halo population and from foreground/background objects. Then, we apply a Sobel filter to detect a sharp transition at $\oi=24.18 \pm 0.08$.  We estimate the TRGB color as $\ogi=1.63 \pm 0.08$ and apply equation (1) of \citet{2019ApJ...884..128O} to calculate $\rm{M}_{\textrm{i,TRGB}}$ as $-3.62 \pm 0.10$. Therefore, the distance modulus to SNFC-dw1 is $\mm = 27.79 \pm 0.12$, corresponding to D $= 3.62 \pm 0.2$ Mpc. To be able to directly compare, we also estimate the TRGB distance of NGC\,253 using our dataset. In this case, we use stars lying in the elliptical annulus covering $3-4\times\rr$ and correct for contamination of foreground/background objects using the same reference field but scaled up to cover the same area within the elliptical annulus. We find $\mm = 27.72 \pm 0.14$, or D $= 3.50 \pm 0.22$ Mpc, which is in excellent agreement with \citet{2011ApJS..195...18R}.   This places SNFC-dw1 at a 3D radius of $\sim146$~kpc on the far side of NGC\,253 but the uncertainties are sufficiently large to make it very tentative.     

\subsection{Luminosity, Structure and Metallicity}
The total luminosity and color of SNFC-dw1 are estimated using the flux of metal-poor RGB stars within the solid circle of the right panel in Figure \ref{fig: map}.  The photometric completeness of a given star is inferred from comparing its magnitude and color with the interpolated completeness curve.  The flux is then scaled to account for stars fainter than the RGB selection box using the best-fit theoretical isochrone.  The contaminant contribution is calculated by applying the same approach to the reference field, and then subtracted.  We find $\mg = -11.15 \pm 0.2$ and $\ogi = 1.35 \pm 0.3$, which corresponds to $\mv=-11.7 \pm 0.2$ using the transformations in \citet{2018ApJ...853...29K}. 
Adopting equation (8) of \citet{2011MNRAS.418.1587T}, we calculate the stellar mass of SNFC-dw1 to be $\sim 1.25\times10^7~\msun$.

\begin{figure}
\begin{center}
\includegraphics[width=230pt]{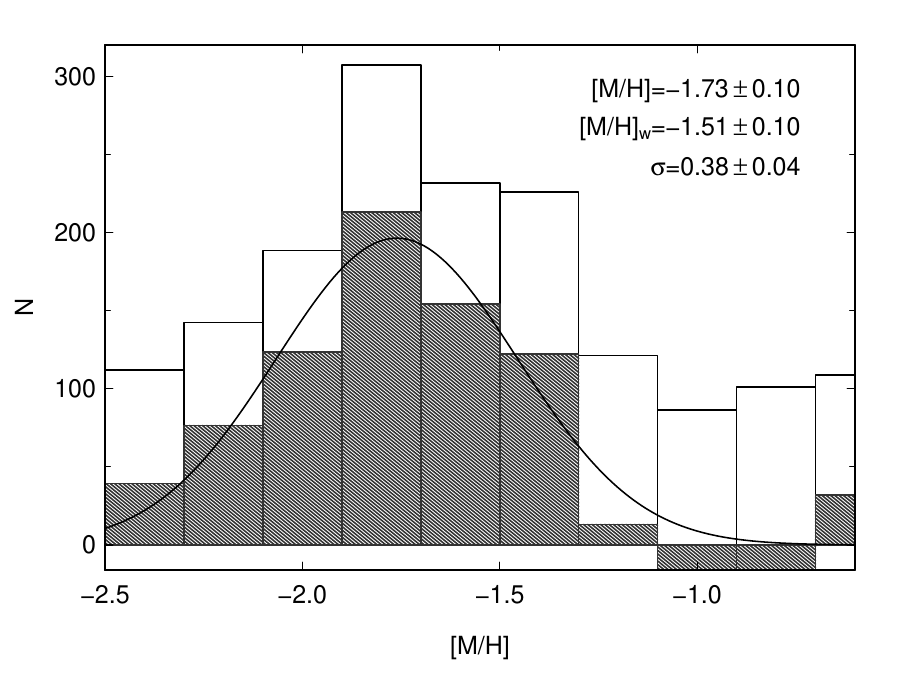}
\caption{Metallicity distribution of RGB stars derived using the $\ogi$ color. The open histogram shows the completeness-corrected MDF, and the shaded histogram illustrates the contaminant-subtracted MDF using the reference field. The solid line shows a Gaussian fit to the contaminant-subtracted MDF.}
\label{fig: mdf}
\end{center}
\end{figure}

Figure \ref{fig: mdf} shows the metallicity distribution function (MDF) of SNFC-dw1 derived from the completeness-corrected color distribution of bright RGB stars. We use RGB stars of $\oi <25.0$ to reduce the uncertainties and to keep the wide spacing between isochrones in the $\ogi$ color.  RGB stars are selected within the solid circle of the right panel of Figure \ref{fig: map} and within the solid polygon of Figure \ref{fig: 4cmds}.  The metallicity is derived from the linear interpolation of isochrones of $\gmh = -2.2$ to $-0.7$ with a fixed age of 10 Gyr and extrapolation with the spline function up to $\gmh = -2.5$ in the same manner as done in \citet{2023ApJ...952...77O}.  The completeness is corrected as before according to the color and magnitude of each star.  The same estimation is done for the reference field in order to derive the contaminant contribution to the MDF arising from the NGC\,253 halo and background/foreground objects.   The contaminant-corrected MDF is shown as the shaded histogram.  The uncertainty is estimated by performing a Monte Carlo simulation with N $= 2000$ for each RGB star.  A star is randomly re-sampled from a Gaussian distribution with a width equal to the photometric error, and the metallicity is redetermined. The standard deviation of the resulting metallicity distribution is adopted as the uncertainty for each RGB star.  The estimated uncertainties increase from 0.03 to 0.14 dex with decreasing metallicity from $\gmh = -0.4$ to $-2.4$ due to the narrow range in RGB color between metal-poor isochrones. The weighted mean metallicity corrected for this effect is shown as the error-weighted mean value $\langle\gmh_{w}\rangle = -1.51 \pm 0.1$ dex.  The overall appearance of the MDF is well-described by a single Gaussian shown as the solid line in Figure \ref{fig: mdf}. 
The metallicity dispersion is significant ($\sigma_{\gmh}=0.38$) compared to the uncertainty (0.04) on the fit. The sudden drop on the metal-rich side ($\gmh > -1.3 $) of the MDF likely reflects an over-subtraction due to small number statistics in this metallicity range, in which the NGC\,253 halo dominates the contaminant signal.    

\begin{deluxetable}{lc}
\tabletypesize{\scriptsize}
\tablecolumns{2}
\tablewidth{0pt}
\tablecaption{The properties of NGC253-SNFC-dw1\label{tbl: str}}
\tablehead{
\colhead{Parameter} \hspace{100pt} & \colhead{Value}\hspace{0pt}}
\startdata  
R.A.(J2000)    & $00^{h}48^{m}39^{s}.68$ \\
Dec. (J2000)   & $-26\arcdeg33\arcmin48\arcsec.7$ \\
Distance       & $3.62 \pm 0.20$ Mpc \\
$\mm$          & $27.79 \pm 0.12 $ \\
P.A.$^{(a)}$         & $105\arcdeg \pm 10\arcdeg $\\
$\epsilon^{(b)}$   & $0.06  \pm 0.04 $ \\
$\mg^{(c)}$        & $-11.15 \pm 0.2$ \\
$\ogi^{(c)}$       & $1.35 \pm 0.3$ \\
$\mv^{(c)}$        & $-11.7 \pm 0.2$ \\
$n^{(d)}$            & $0.45 \pm 0.04$ \\
$\rh^{(e)}$        & $3.20\arcmin \pm 0.16\arcmin$  \\
$\rh^{(e)}$        & $3.37 \pm 0.36$  kpc\\
$\langle\gmh_{w}\rangle^{(c)}$  & $-1.51 \pm 0.10$ dex\\
$\sigma_{\gmh}^{(c)}$ & $0.38 \pm 0.04  $ dex \\
$\langle\mu_{g}\rangle (r<\rh)^{(f)}$       & $\sim 30.1$ mag arcsec$^{-2}$ \\
$\textrm{M}_*^{(g)}$  & $\sim 1.25\times10^7 \msun$
\enddata
\tablecomments{(a) Position angle from north to east. 
(b) Ellipticity $\epsilon = 1 - b/a$ where $b/a$ is the axis ratio.  
(c) Measured using stars within a circular radius of 9 arcmin at the SNFC-dw1. 
(d) S\'{e}rsic index.  
(e) Half-light radius of the S\'{e}rsic profile. 
(f) Average surface brightness within the half-light radius in g-band.
(g) The stellar mass calculated using the inferred total magnitude.}
\end{deluxetable}

Although the extended regions of SNFC-dw1 are very irregular, the core structure seen in the lower-middle panel of Figure \ref{fig: map} is well-defined and so we proceed to estimate its structural properties.  We compute the values using the density-weighted zeroth, first and second moments of the metal-poor RGB spatial distribution and list them in Table \ref{tbl: str}.  The estimated shape shown as a dashed line in Figure \ref{fig: map}  seems to be more circular than the visual impression of the density map, probably due to the existence of extended components toward the north.  We also fit the standard S\'{e}rsic profile via least-squares minimization to the completeness-corrected metal-poor RGB star counts. The radial profile is constructed by calculating the average number density of RGB stars in a series of elliptical annuli, defined using the derived structural parameters. We take the Poisson uncertainties in the RGB counts as the uncertainty. Initially, the contaminant contribution is visually estimated as the average number density at large radius ($>10\arcmin$) and we fit the profile and estimate the half-light radius ($\rh$).  We then iterate on the estimation of the contaminant level, this time using the average number density between 4 and  $6\times\rh$. Finally, the profile is refit with this contamination level.  The resulting $\rh=3.37 \pm 0.36$ kpc is very large for a dwarf galaxy and, in combination with the relatively low luminosity, yields an extremely faint average surface brightness of $\mu_{g} \sim 30.1$ mag arcsec$^{-2}$ within $\rh$. 

\begin{figure}
\begin{center}
\includegraphics[width=250pt]{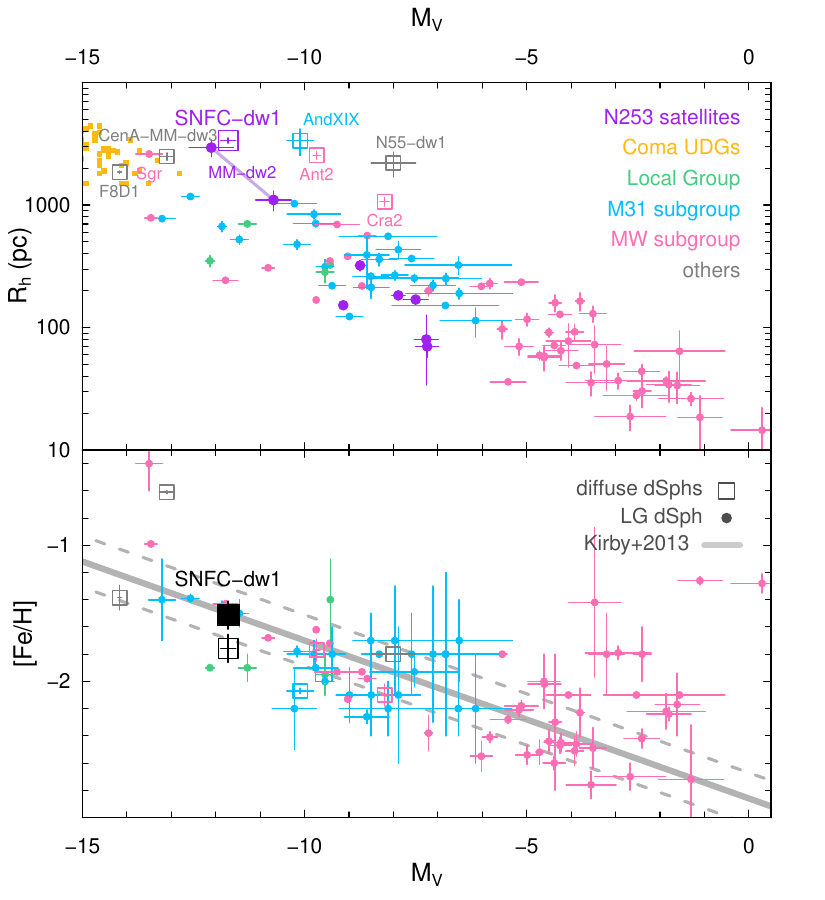}
\caption{{\it Upper:} The luminosity-size relation. The purple symbols are known NGC\,253 satellites \citep{2021A&A...652A..48M,2022ApJ...926...77M}. The yellow squares are UDGs in the Coma cluster \citep{2015ApJ...798L..45V}. The filled circles show LG dSphs from \citet[][updated in January 2021]{2012AJ....144....4M}. The open squares with names are particularly diffuse dSphs  \citep{2016ApJ...816L...5T, 2019MNRAS.488.2743T, 2019ApJ...872...80C, 2020MNRAS.491.3496C, 2021ApJ...921...32J,2022ApJ...938..101S, 2023MNRAS.518.2497Z}.
{\it Lower:}  The luminosity-metallicity relation of nearby dSphs. The filled and open black squares represent SNFC-dw1 with $\afe = 0.0$ and 0.32 dex, respectively. The solid and dotted lines show the \citet{2013ApJ...779..102K} relation with the $1\sigma$ deviation. }
\label{fig: MV}
\end{center}
\end{figure}

\section{Discussion and Summary}
Using data from the SNFC survey being conducted with the Subaru Telescope, we have uncovered a new dwarf satellite of the nearby spiral NGC\,253. The system lies at a projected separation of 83.4 kpc from the center of NGC\,253.  We have used the TRGB method to derive consistent distances to SNFC-dw1 and NGC\,253, finding them to lie at $3.62\pm 0.2$ Mpc and $3.5\pm0.2$ Mpc, respectively. This places SNFC-dw1 
well within its estimated virial radius of 330~kpc \citep{2021ApJ...918...88M}, confirming it is a bound satellite.  

We find SNFC-dw1 to be dominated by an old metal-poor population; additionally,  there is tentative evidence for an associated TP-AGB population.   As such stars have ages of typically $0.6-2$~Gyr \citep{2017ApJ...835...77M},  their presence suggests that the system was star-forming until relatively recently. 
Although SNFC-dw1 has a comparable luminosity, stellar mass, and metallicity to bright classical Galactic dSphs, it is extremely diffuse ($\mu_{g} \sim 30.1$ mag arcsec$^{-2}$) -- a direct consequence of its abnormally large size with $\rh = 3.37 \pm 0.36$ kpc.  Indeed, star counts can be traced over more than 10 kpc and reveal a very asymmetric structure in the peripheral regions.  Notably, SNFC-dw1 has a size comparable to some of the most diffuse galaxies currently known, including the puzzling Local Group systems Crater\,2 ($\rh=1.1$~kpc), Antlia\,2 ($\rh=2.5$~kpc) and Andromeda\,XIX ($\rh=3.4$~kpc) \citep{2016MNRAS.459.2370T, 2021ApJ...921...32J, 2022ApJ...938..101S}.

Figure \ref{fig: MV} compares SNFC-dw1 to other dwarf galaxies in terms of scaling relationships. These plots include LG dSphs \citep[][updated in January 2021]{2012AJ....144....4M}, Ultra-diffuse galaxies (UDGs) in the Coma Cluster \citep{2015ApJ...798L..45V}, some of the known NGC\,253 satellites \citep{2016ApJ...816L...5T,2021A&A...652A..48M,2022ApJ...926...77M} and the particularly diffuse dSphs (open squares) Antlia\,2 \citep{2019MNRAS.488.2743T, 2021ApJ...921...32J,2022ApJ...926...78V}, Andromeda\,XIX \citep{2020MNRAS.491.3496C}, the M81 satellite F8D1 \citep{2023MNRAS.518.2497Z}, the Centauras A satellite CenA-MM-dw1 \citep{2019ApJ...872...80C}, and the recently-discovered NGC\,55-dw1 \citep{2024ApJ...961..126M}. The upper panel of Figure \ref{fig: MV} shows the systems on the size-luminosity plane. Only six systems have a similarly large size ($\rh \gtrsim 1.5$~kpc) to SNFC-dw1 -- Andromeda\,XIX and Antlia\,2 and NGC\,55-dw1,  which are $\gtrsim 2$ magnitudes fainter, and the Sagittarius (Sgr) dSph, CenA-MM-dw3 and F8D1, which are $\gtrsim 2$ magnitudes brighter.  While both Sgr, CenA-MM-dw3 and F8D1 show direct evidence for tidal disruption \citep{2016ApJ...823...19C, 2003ApJ...599.1082M, 2023MNRAS.518.2497Z}, the evidence in Antlia\,2 and Andromeda\,XIX has thus far only been circumstantial \citep[e.g.][]{2006ApJ...642L.137B,2019MNRAS.488.2743T,2020MNRAS.491.3496C,2021ApJ...921...32J}.  The situation is also not clear for NGC\,55-dw1 which, in addition to lacking clear tidal features, does not yet have a directly measured distance \citep{2024ApJ...961..126M}.  Amongst the NGC\,253 satellites, MM-dw2 is most similar to SNFC-dw1 and it exhibits a highly elongated shape (see Figure \ref{fig: map}). However, the properties of MM-dw2 estimated from two independent studies are rather different -- $\mv = -12.1$ and $\rh=2.94$ kpc using Magellan/Megacam \citep{2016ApJ...816L...5T} and $\mv = -10.7$ and $\rh=1.1$ kpc using Subaru/Suprime-Cam \citep{2016MNRAS.457L.103R}. We plot both values connected with a purple solid line in the upper panel of Figure \ref{fig: MV}.  The prominent tidal extensions emanating from SNFC-dw1 provide unambiguous evidence of tidal disruption, demonstrating that this is a natural pathway to produce such extremely diffuse and extended galaxies.  This conclusion is further supported by integrated light analyses of two more distant dwarfs with similarly large sizes -- NGC4449B associated with the dwarf starburst galaxy NGC 4449 \citep{2012Natur.482..192R, 2012ApJ...748L..24M} and HCC-087 in the Hydra I galaxy cluster \citep{2012ApJ...755L..13K}, both of which are also known to exhibit S-shaped tidal features. 

The lower panel of Figure \ref{fig: MV} shows the luminosity-metallicity scaling relation (LZR) defined by local dwarfs.  To place SNFC-dw1 on this plot, we converted the global metallicity $\gmh$ to $\feh$ using equation (3) of \citet{1993ApJ...414..580S} assuming two cases, $\afe=0.0$ and 0.32 dex, the latter inferred from recent spectroscopic studies of UDGs \citep{2023MNRAS.526.4735F}.  
Broadly speaking,  SNFC-dw1 obeys the same LZR as LG dSphs \citep{2013ApJ...779..102K}.
If we assume an $\alpha$-enhanced abundance, it is somewhat more metal-poor than expected for a galaxy of the same luminosity but it still lies within $\sim 1\sigma$ of the relation, suggesting that it has not yet lost a significant fraction of its original stellar mass.   Whether or not this finding is at odds with the highly disrupted appearance of SNFC-dw1 is unclear.  A measurement of the velocity dispersion would be extremely valuable for constraining the degree of tidal stripping \citep[e.g.][]{2022MNRAS.512.5247B} but such observations would be very challenging with current facilities given the faintness of SNFC-dw1's stars. 

It is worth noting that SNFC-dw1 lies in an area of the sky that has been the target of multiple dwarf galaxy searches in recent years, some of which have used integrated light \citep{2021A&A...652A..48M, 2022ApJ...926...77M} while others have used resolved stars \citep{2022ApJ...933...47C}. Although it is the second most luminous of the six satellites projected within 150~kpc of NGC\,253 \citep{2024arXiv240114457M}, and comparable to the brightest classical dSph of the Milky Way and Andromeda, SNFC-dw1 has remained unearthed until now due to its extremely diffuse nature.  This emphasizes the importance of surface brightness limiting depth in satellite searches, suggesting that even bright-end incompleteness can be an issue at the typical depths of most present-day studies. Fortunately, with the recent launch of ESA's Euclid satellite, the start of the Legacy Survey of Space and Time with the Vera C. Rubin Observatory in 2025 and the Nancy Grace Roman Space Telescope launch in 2027, the era of very sensitive ($\mu > 30$ mag arcsec$^{-2}$) satellite searches around large numbers of Local Volume galaxies will soon be upon us.


\section{Acknowledgments}
We thank Bur\c{c}in Mutlu-Pakdil,  David Sand and the anonymous referee for helpful comments. 

SO acknowledges support in part from JSPS Grant-in-Aid for Scientific Research (18H05875, 20K04031, 20H05855). AMNF and RZ are grateful for support from the UK STFC via grants ST/Y001281/1 and ST/V000594/1. JK acknowledges support from NSF through grants AST-1812847 and AST-2006600. MGL was supported by the National Research Foundation grant funded by the Korean Government (NRF-2019R1A2C2084019). 

We are grateful to all the staff at Subaru Telescope and the HSC team. This research is based on data collected at the Subaru Telescope, operated by the National Astronomical Observatory of Japan. We are honored and grateful for the opportunity to observe the Universe from Maunakea, which has cultural, historical, and natural significance in Hawaii. 

This paper makes use of software developed for Vera C. Rubin Observatory. We thank the Rubin Observatory for making their code available as free software at http://pipelines.lsst.io/.  The Pan-STARRS1 Surveys (PS1) and the PS1 public science archive have been made possible through contributions by the Institute for Astronomy, the University of Hawaii, the Pan-STARRS Project Office, the Max-Planck Society and its participating institutes, the Max Planck Institute for Astronomy, Heidelberg and the Max Planck Institute for Extraterrestrial Physics, Garching, The Johns Hopkins University, Durham University, the University of Edinburgh, the Queen's University Belfast, the Harvard-Smithsonian Center for Astrophysics, the Las Cumbres Observatory Global Telescope Network Incorporated, the National Central University of Taiwan, the Space Telescope Science Institute, the National Aeronautics and Space Administration under Grant No. NNX08AR22G issued through the Planetary Science Division of the NASA Science Mission Directorate, the National Science Foundation Grant No. AST-1238877, the University of Maryland, Eotvos Lorand University (ELTE), the Los Alamos National Laboratory, and the Gordon and Betty Moore Foundation. 

For the purpose of open access, the author has applied a Creative Commons Attribution (CC BY) licence to any Author Accepted Manuscript version arising from this submission.


%

\vspace{5mm}
\facilities{Subaru(HSC)}






\bibliography{main}{}
\bibliographystyle{aasjournal}



\end{document}